\documentclass[aps,prd,nofootinbib,twocolumn,amsmath,amssymb,longbibliography]{revtex4-1}

\usepackage{amsmath,amssymb,amsfonts,latexsym,cancel}

\usepackage{mathtools}
\usepackage{graphicx}
\usepackage{color}
\usepackage{amsbsy}
\usepackage{hyperref}
\usepackage{tikz}

\usepackage{seqsplit}

\usepackage[caption=false]{subfig}
\usepackage{float}

\usepackage{comment}

\newcommand{\be}{\begin{equation}}
\newcommand{\beq}{\begin{equation}}
\newcommand{\ee}{\end{equation}}

\def\bea {\begin{eqnarray}}
\def\eea {\end{eqnarray}}

\def\f{\frac}

\def\lp{\ell_{\rm Pl}}

\def\dd{{\rm d}}

\begin{document}

\title{Dynamical homogenization in effective loop quantum cosmology}

\author{Edward Wilson-Ewing} \email{edward.wilson-ewing@unb.ca}
\affiliation{Department of Mathematics and Statistics, University of New Brunswick, \\
Fredericton, NB, Canada E3B 5A3}

\begin{abstract}

The effective dynamics of loop quantum gravity for marginally bound Lema\^itre-Tolman-Bondi spacetimes predict that the big-bang singularity is resolved and replaced by a cosmic bounce. Numerics show that these effective dynamics also homogenize small regions soon after the bounce when inhomogeneities before the bounce are sufficiently large. These homogeneous regions typically have a width of a few Planck lengths where relative perturbations in the energy density remain less than 15\%. If the bounce is followed by an inflationary period, the homogeneous region can reach cosmic scales and the amplitude of relative perturbations can be suppressed to a value compatible with observations.

\end{abstract}

\maketitle

\section{Introduction}

The small amplitude of the temperature fluctuations $\delta T / T \sim 10^{-5}$ observed in the cosmic microwave background (CMB) \cite{Planck:2018nkj}, combined with the assumption of the Copernican principle that our location in the universe is a typical one, lead to the conclusion that the early universe was extremely homogeneous.  This homogeneity provides initial conditions for the later evolution of the universe that matches observations, but begs the question as to why the early universe is so highly homogeneous in the first place. Further, given the expectation that gravity will dominate the dynamics at the largest scales and, being attractive, can be expected to cause inhomogeneities to grow, it seems an even higher degree of homogeneity is required at times well before the formation of the CMB.

This is a fine-tuning problem if there is no dynamical mechanism for homogenizing, or smoothing, the early pre-CMB universe. Importantly, there do exist some such proposals, notably through inflation \cite{Starobinsky:1982mr, Linde:1985ub, Jensen:1986nf, East:2015ggf, Clough:2016ymm, Kleban:2016sqm, Creminelli:2020zvc, Wang:2021hzv} (although see \cite{Garfinkle:2023vzf}) or ekpyrosis \cite{Garfinkle:2008ei, Cook:2020oaj}; this letter proposes another possibility where homogenization occurs during a cosmic bounce.

Inflation can alleviate the homogeneity problem in two ways. First, if there is a small region that is moderately homogeneous, then inflationary expansion can blow this small region up to cosmological scales and further homogenize it \cite{Linde:1985ub}; and second, under some assumptions, it has been shown that large homogeneous regions will necessarily have formed at sufficiently late times in an inflationary spacetime \cite{Starobinsky:1982mr, Jensen:1986nf, Kleban:2016sqm, Creminelli:2020zvc, Wang:2021hzv}. Note however that in the first case, it is necessary to already have small moderately homogeneous regions before inflation, while for the second family of results extremely long periods of inflation may be required for sufficient smoothing. As shown below, the cosmic bounce of loop quantum cosmology can generate the small and moderately homogeneous regions needed for the first scenario.

A phase of ekpyrotic contraction is an alternative mechanism that can efficiently smooth out inhomogeneities. This smoothing is successful even if the initial departures from homogeneity are of order one, and it can homogenize large regions of space \cite{Garfinkle:2008ei, Cook:2020oaj}. The drawback of ekpyrosis is that multiple fields are needed to generate (nearly) scale invariant scalar perturbations \cite{Finelli:2002we, DiMarco:2002eb, Lehners:2007ac} and the perturbations are sensitive to anisotropies \cite{Agullo:2022klq}, which introduces a separate fine-tuning problem (this is in contrast to the background ekpyrotic dynamics which are robust to the presence of anisotropies).

Another possibility is that quantum gravity could help to alleviate the inhomogeneity problem, as was first suggested based on the observation that in loop quantum cosmology, the big-bang singularity is resolved and replaced by a non-singular bounce; heuristically, this can be interpreted as quantum gravity becoming repulsive at Planckian spacetime curvature. Could this effective repulsion perhaps also smooth out inhomogeneities during the cosmic bounce \cite{Agullo:2012sh}?

Although an intriguing idea, perturbative calculations give no sign of such an effect. In classical general relativity, linearized cosmological perturbations grow in contracting spacetimes if the equation of state $w = p/\rho$ relating the energy density $\rho$ and pressure $p$ of the matter field satisfies $-1 < w \le 1$. Assuming the strong energy condition $w>-1$, the remaining possibility is ekpyrotic fields satisfying $w>1$ for which the amplitude of the cosmological perturbations remains constant. In this perturbative analysis, therefore, a contracting pre-bounce phase will not homogenize the spacetime as inhomogeneities either grow or at best have a constant amplitude.

What about the bounce itself? A simple model-independent approach is to approximate the bounce as a discontinuous change in the Hubble rate $H$ from $H=-H_\star$ to $H=H_\star$ and require that cosmological perturbations satisfy appropriate matching conditions \cite{Deruelle:1995kd}; in this calculation the amplitude of the perturbations remains of the same order of magnitude as before the bounce \cite{Brandenberger:2001bs, Hwang:2001ga, Copeland:2006tn}. Alternatively, it is possible to consider specific theories that predict a cosmic bounce (including loop quantum cosmology) and explicitly calculate the evolution of the perturbations through the bounce: the result is the same, with no suppression of the amplitude of cosmological perturbations across the bounce; for reviews of these calculations in general bouncing cosmological scenarios as well as specifically in loop quantum cosmology, see, respectively, \cite{Brandenberger:2016vhg, Wilson-Ewing:2015sfx}.

Nonetheless, these previous analyses are based on the truncation of inhomogeneities to the approximation of linear perturbation theory. As shall be shown next, a non-perturbative analysis of this problem in Lema\^itre-Tolman-Bondi (LTB) spacetimes gives the opposite result: in effective loop quantum gravity, an inhomogeneous cosmic bounce can indeed homogenize small regions. Although these regions are small, they can later be expanded to cosmological scales, and also further homogenized, by inflation \cite{Linde:1985ub}.

\section{Homogenization in Lema\^itre-Tolman-Bondi Spacetimes}
\label{s.ltb}

The Lema\^itre-Tolman-Bondi (LTB) spacetime, corresponding to spherically symmetric dust, is an exact solution to the Einstein equations that allows arbitrarily large inhomogeneities in the radial direction. It is convenient to write the metric in generalized Painlev\'e-Gullstrand coordinates,
\beq \label{metric}
\dd s^2 = - \dd t^2 + \f{1}{1 + \mathcal{E}} (\dd r + N^r \dd t)^2 + r^2 \dd\Omega^2,
\ee
where the shift vector $N^r$ depends on the radius $r$ and time $t$ coordinates, and $\dd\Omega^2 = \dd\theta^2 + \sin^2 \theta \, \dd\phi^2$ is the metric on the unit 2-sphere. For simplicity, I will focus on the $\mathcal{E} = 0$ marginally bound configurations.

As a first step in exploring loop quantum gravity (LQG) effects, there are effective equations that include corrections due to the discrete nature of quantum geometry as predicted in LQG, but that neglect quantum fluctuations; specifically, the discrete nature of quantum geometry is captured by evaluating holonomies along paths of a minimal Planckian length. Such effective equations were first derived in the context of homogeneous cosmology \cite{Ashtekar:2006wn, Taveras:2008ke}, and for sharply-peaked quantum states they provide an excellent approximation to the dynamics of expectation values, so long as the volume of the spatial region considered is much larger than the Planck volume \cite{Rovelli:2013zaa, Bojowald:2015fla}.

Similarly, effective equations can also be derived for the LTB spacetime \cite{Kelly:2020lec, Husain:2021ojz, Husain:2022gwp, Giesel:2023hys}, that are expected to be a good approximation to the quantum dynamics at large length scales, but quantum fluctuations could become important at shorter scales. As a starting point, here I focus on the effective dynamics, and leave the inclusion of quantum fluctuations for future work.

For marginally bound LTB spacetimes expressed in generalized Painlev\'e-Gullstrand coordinates, there is a single (local) degree of freedom corresponding to the angular component $b$ of the Ashtekar-Barbero connection. In the effective LQG theory \cite{Kelly:2020lec, Husain:2021ojz, Husain:2022gwp, Giesel:2023hys},
\beq \label{eom-b}
\partial_t b = - \f{1}{2 \Delta r} \partial_r \left( r^3 \sin^2 \f{\sqrt\Delta b}{r} \right),
\ee
where $\Delta \sim \lp^2$ is the smallest non-zero eigenvalue of the area operator in LQG, and the Barbero-Immirzi parameter is set to 1. The velocity $v_b$ of the field (i.e., the prefactor to $\partial_r b$ after expanding the equation of motion) is
\beq \label{vb}
v_b = - \f{r}{\sqrt\Delta} \sin \f{\sqrt\Delta b}{r} \cos \f{\sqrt\Delta b}{r}.
\ee

Then, $N^r = v_b$ \cite{Gambini:2020nsf, Kelly:2020uwj, Husain:2022gwp, Giesel:2023hys}, and $b$ also determines the dust energy density $\rho$ by \cite{Kelly:2020lec, Husain:2021ojz, Husain:2022gwp}
\beq \label{rho-b}
\rho = \f{\rho_c}{3 r^2} \, \partial_r \left( r^3 \sin^2 \f{\sqrt\Delta b}{r} \right),
\ee
where $\rho_c = 3 / 8 \pi G \Delta$ is the critical energy density of loop quantum cosmology.

In homogeneous cosmology, the big-bang singularity is replaced by a non-singular bounce that occurs at $v_b(t) = 0$ \cite{Ashtekar:2006wn}. As shown below, the dynamics are to some extent qualitatively similar for marginally bound LTB spacetimes with the difference that the bounce is inhomogeneous since $v_b$ is now a function also of $r$, and the bounce at each radius occurs at different times, when $v_b(r,t) = 0$.

The fact that the bounce can occur at different times at different radii has important consequences. In particular, if there is a region $r \in [r_1, r_2]$ with a large over-density, it may bounce before the region $r < r_1$ lying immediately inside. Then, the outer region that bounces first starts to expand with $v_b^{(out)} > 0$ while the inner region continues to contract with $v_b^{(in)} < 0$. This causes a rarefaction wave (located around $v_b=0$) to form between the two regions, with the rarefaction wave moving inwards, following the inner region that continues to contract. In the meantime, the outer region is expanding, and the new intermediate region located between the ingoing rarefaction wave and the outermost part of the now-expanding outer region will be quite homogeneous, because all of these points will have had $v_b=0$ nearly simultaneously, implying from \eqref{vb} and \eqref{rho-b} that $\rho$ is nearly constant in this region.

Note that these regions are nearly, but not exactly, homogeneous. This is because the rarefaction wave (where $v_b=0$, and therefore the location where the bounce occurs) moves inwards, and as a result at larger $r$ the bounce happens at slightly earlier times, giving $\rho$ more time to dilute during the post-bounce expanding phase: $\rho$ will then be monotonically decreasing with $r$ in nearly homogeneous regions, as confirmed by the numerics (discussed below) that clearly show this (small) slope.

The size of homogeneous regions can be increased by a longer-lived rarefaction wave (that vanishes when the less dense inner region bounces), caused by larger differences in $\rho$ in the inner and outer regions: ironically, greater inhomogeneities pre-bounce lead to larger homogeneous regions post-bounce.

To solve the dynamics numerically, it is convenient to introduce $B = rb$, so the dynamics are given by a conservation law $\partial_t B + \partial_r f(B; r) = 0$ with $f = (r^3/2\Delta) \cdot \sin^2 (\sqrt\Delta B / r^2)$. Then, the Godunov algorithm for non-linear wave equations can be used \cite{Leveque}; this has been done for the LQG effective dynamics of LTB spacetimes for configurations of $\rho$ that have compact support \cite{Husain:2021ojz, Husain:2022gwp, Cipriani:2024nhx}. For the cosmological configurations of interest here, the key difference is to the outer boundary condition: the inner boundary condition $B(r=0)=0$ remains unchanged (due to $B = rb$ vanishing at $r=0$), but the outer boundary condition at $r=r_{max}$ is fixed by assuming a cosmological exterior with homogeneous energy density implying $B(r=r_{max}) \propto r^2$. For a discretization in $r$ of constant width $\delta r$, this fixes $B(r_{max})$ at each time step through $B(r_{max}) = r_{max}^2 B(r_{max} - \delta r) / (r_{max} - \delta r)^2$.

\begin{figure}[t]
    \centering
    \subfloat[Initial conditions.]{\includegraphics[width=0.4\textwidth]{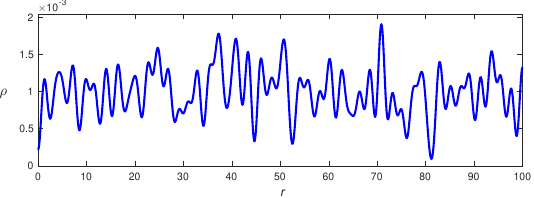} \label{fig:framesA}} \\
    \subfloat[During contraction, $t=3.00$.]{\includegraphics[width=0.4\textwidth]{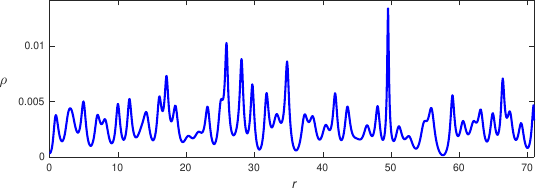} \label{fig:framesB}} \\
    \subfloat[During the bounce, $t=7.05$.]{\includegraphics[width=0.4\textwidth]{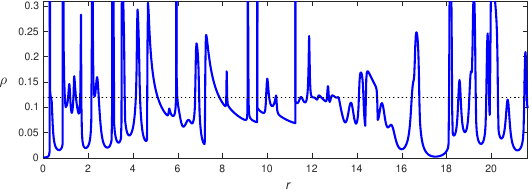} \label{fig:framesC}} \\
    \subfloat[Post-bounce expansion, $t=8.40$.]{\includegraphics[width=0.4\textwidth]{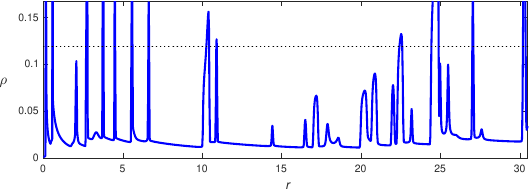} \label{fig:framesD}}
    \caption{\footnotesize These frames from a representative example show $\rho(r)$ at four instants, using units where $\hbar = \Delta = 1$. The top frame shows the initial conditions, the second shows a later time during contraction, while the third frame shows part of the bounce, with the region located approximately between $r\approx 7$ and $r \approx 11$ having already bounced and the region located approximately between $r\approx 11$ and $r \approx 13$ in the process of bouncing, and 
the fourth frame shows the time by which $R$ expanded by a factor of 1.5 after the bounce. The critical energy density $\rho_c$ is shown as a dotted line in the bottom two frames. The formation of small homogeneous regions can be seen in the last frame: there are four regions with a radial extent of at least $\sqrt\Delta$ where $|\partial_r \rho| / \rho < 0.15$, located around $r = 9$, $r = 12$, $r = 15$, and $r = 29$. Note that the range of $r$ differs for each frame, in order to follow the same spacetime region as it initially contracts and later expands.}
    \label{fig:frames}
\end{figure}

Initial conditions are given for $r \in [0, r_{max}]$ corresponding to a contracting spacetime with inhomogeneities in $\rho$ of order 1, and the Godunov algorithm is used to solve the dynamics numerically. The initial conditions are given in terms of $\rho$ as
\beq
\rho(t_i) = \rho_{avg} + \sum_{n=1}^{m} \alpha_n \cos \left( \f{n \pi r}{r_{max}} \right),
\ee
where $\rho_{avg} > 0$, the $\alpha_n$ are randomly selected from a uniform distribution in the same interval $[-A,A]$, with $A$ chosen so perturbations are large while keeping $\rho > 0$. The Fourier expansion is used only for setting initial conditions, and the sum over $n$ ends at $m = r_{max} / \sqrt\Delta$, corresponding to a Planck-scale cutoff on the Fourier expansion for $\rho(t_i)$. Note that $\sin(n \pi r / r_{max})$ terms are excluded from the series to ensure that $\rho$ is smooth at the origin. Then, $\rho(t_i)$ is used to determine $B(t_i)$ by inverting \eqref{rho-b}, after which the Godunov algorithm is used to evolve $B$. The lattice spacing $\delta r$ in the radial direction is fixed, while the time steps $\delta t$ are determined dynamically using the Courant-Friedrichs-Lewy condition \cite{Leveque}. Once $b=B/r$ is known, the energy density $\rho$ at later times is calculated from \eqref{rho-b}. The code is available online \cite{code}.

The numerical solutions depend on the initial conditions, but there are certain key features that appear generically: (i) there is a bounce in the cosmological dynamics also in the presence of large inhomogeneities, like in homogeneous cosmology \cite{Ashtekar:2006wn}; (ii) shock waves form during contraction, as is guaranteed if inhomogeneities are sufficiently large \cite{Hellaby:1985zz, Fazzini:2023ova}; and (iii) small homogeneous regions emerge during the bounce. Due to the formation of shocks, it is necessary to allow for weak solutions, as is done by the Godunov algorithm; for more on weak solutions in effective LQG see \cite{Husain:2021ojz, Husain:2022gwp, Fazzini:2023ova}. As explained above, the homogeneous regions form due to an outer over-dense region bouncing before a neighbouring inner under-dense region, causing the formation of a rarefaction wave moving inwards so that a continuous interval of $r$ bounces nearly simultaneously.

These dynamics are not invariant under time reversal. This is a general feature of weak solutions \cite{Leveque}, and in particular the formation of the rarefaction wave breaks time-reversal symmetry. This is one reason why the formation of homogeneous regions can be generic, and does not require a certain class of initial conditions.

To quantitatively characterize the homogeneous regions formed during the bounce, a reasonable definition that I will use here is that they are intervals in $r$ of width greater than $\sqrt\Delta$ where for each $r$ in that interval, $|\partial_r \rho| / \rho < 0.15$ (results are similar for other possible definitions of homogeneous regions).

A typical example, with $r_{max} = 100, \delta r = 10^{-4}, \rho_{avg} = 10^{-3}, A = 0.08$ (using units where $c = \hbar = \Delta = 1$), is shown in Fig.~\ref{fig:frames}.
Note that in terms of the coordinate $r$, during contraction the dust will initially move inwards, and then outwards after the bounce; the results are depicted using the range for $r$ from the origin to $R(t)$, the outer edge of the inhomogeneous region (that is initially $R(t_i) = r_{max}$). For the example shown in Fig.~\ref{fig:frames}, $R$ decreased by a factor of $\sim 4.9$ during contraction, and increased by a factor of $\sim 1.5$ after the bounce, at which time the numerics were stopped. For this run, at the final time there are four homogeneous regions where $|\partial_r \rho / \rho| < 0.15$ with widths of 1.93, 3.04, 1.54, and 2.12, and located around $r=9, 12, 15$ and $29$ respectively; note that these cover over a quarter of the range of the numeric results. It is worth pointing out that the gradient in $\rho$ is for the most part due to a linear and quadratic dependence on $r$ in the homogeneous regions, with the linear best fit accounting for approximately one third of $(\partial_r \rho)/\rho$, and the quadratic best fit accounting for another third in each of the four homogeneous regions.

As can be seen in Fig.~\ref{fig:framesD}, there are typically no bumps or other types of features in the homogenized regions, except for a small decrease in $\rho$, due to the outer region bouncing first, and therefore expanding and diluting more, as explained above.

On its own, a smoothing to $|\partial_r \rho / \rho| < 0.15$ over a length scale of a few $\sqrt\Delta$ is not sufficient to explain perturbations of amplitude $\sim 10^{-5}$ at cosmological scales, but if this is combined with a later sufficiently long-lasting era of inflation then the size of the smoothed region will be exponentially enlarged, and also further homogenized, by an amount depending on the number of $e$-folds of inflation. In particular, the amplitude of the relative perturbations over the observable universe can be made arbitrarily small by choosing a sufficiently large number of $e$-folds. In addition, inflation can also strongly isotropize the universe after the LQG bounce, even starting from initially highly anisotropic configurations \cite{Gupt:2013swa}.

Note however that if there is a limited amount of inflation, the remaining small near-linear change in $\rho$ could potentially have an observational impact, for example by introducing a preferred direction in the CMB corresponding to the direction where $\rho$ increases. More work is needed to determine whether this could cause the hemispherical power asymmetry observed in the CMB \cite{Planck:2015igc} (in particular, such a study will require going beyond spherical symmetry as well as including more realistic matter fields); a detailed investigation of this possibility is left for future work.

\section{Discussion}

In conclusion, in marginally bound LTB spacetimes the cosmic bounce of loop quantum cosmology can provide a mechanism to smooth out inhomogeneities in the very early universe, forming small moderately homogeneous regions. These regions can then be blown up to cosmological size and further homogenized by inflation, as argued in \cite{Linde:1985ub}, thereby potentially solving the homogeneity problem of the early universe. Note that in this scenario, the Copernican principle fails at the largest (super-horizon) scales, where there exist significant inhomogeneities.

The physics underlying the homogenization process is not as simple as quantum gravity becoming repulsive when the spacetime curvature is sufficiently large (indeed, some regions are not homogenized during the bounce, contrary to what would be expected if quantum gravity were to become `universally repulsive' at Planck scales). Rather, if inhomogeneities are sufficiently large, then a rarefaction wave will form between an outer sufficiently over-dense region adjacent to an inner sufficiently under-dense region. It is the new intermediate region between the outer region (that bounces first) and the rarefaction wave (that moves inwards following the inner region that contracts for a longer time), that is approximately homogeneous after the bounce.

There are two limitations to the effective dynamics studied here. First, the effective equations of motion are derived in a specific gauge corresponding to the choice of generalized Painlev\'e-Gullstrand coordinates, and for this reason it is not possible to check the covariance of the effective model within this gauge-fixed approach. There has been recent progress in developing explicitly covariant effective models for spherically symmetric black holes \cite{Alonso-Bardaji:2021tvy, Alonso-Bardaji:2023vtl, Duque:2023syb, Zhang:2024khj, Belfaqih:2024vfk, Zhang:2024ney}, and an important next step would be to extend these works to the cosmological context considered here. Second, to allow for the potential formation of shock waves as well as the rarefaction waves that play a central role in the homogenization process described above, it is necessary to allow for weak solutions to the effective dynamics; however, a well known property of weak solutions is that they are not unique \cite{Leveque}. Therefore, a class of solutions must be selected, for example through the choice of a preferred fundamental variable in terms of which the integral equation of motion is expressed. Some input from the microscopic physics underlying the coarse-grained description governed by the integral equation is normally needed for this, which in this case is quantum gravity; for this reason, weak solutions offer a window from the coarse-grained picture provided by the effective dynamics onto the microscopic degrees of freedom of quantum gravity \cite{Fazzini:2023ova}. Here, motivated by loop quantum gravity and following earlier work studying black hole collapse models \cite{Husain:2021ojz, Husain:2022gwp, Cipriani:2024nhx}, the fundamental variable is chosen to be (a component of) the Ashtekar-Barbero connection, but other choices are possible that predict different physics \cite{Fazzini:2025hsf}. A key point though is that the non-uniqueness typically only affects the dynamics of shock waves and rarefaction waves once they have formed, but that the formation of shock waves and rarefaction waves is usually independent of the choice of the weak solution (a rare exception is when there exists a discontinuity in terms of one variable but not another, as can occur in highly symmetric configurations like the Oppenheimer-Snyder collapse model---compare \cite{Kelly:2020lec, Husain:2022gwp, Lewandowski:2022zce, Fazzini:2023scu}---but this possibility is ruled out here due to the large inhomogeneities that guarantee the formation of shell-crossing singularities and therefore of shock waves \cite{Fazzini:2023ova}). Because of this, in this effective model of the LTB spacetime, shock waves and rarefaction waves will generically occur if inhomogeneities are sufficiently large pre-bounce, and generate the homogenization effect studied here. On the other hand, the post-bounce size of the homogenized regions will depend on the specific weak solution---for other choices, it is reasonable to expect that the homogeneous region could be smaller or larger than what is found here; a detailed analysis of this question is left for future work.

Also, note that the specific model considered here, namely the marginally bound Lema\^itre-Tolman-Bondi spacetime with LQG effective dynamics, provides a comparatively simple framework to study homogenization in the early universe. This has the important advantage of making calculations tractable, but it is important in future work to consider richer models to check the robustness of these results.

A first step in this direction is to include non-vanishing pressure---pressure may help strengthen the homogenization process, but this needs to be verified; alternatively, it may decrease inhomogeneities during contraction and suppress the formation of homogeneous regions during the bounce. A more challenging next step is to move beyond spherical symmetry and allow general inhomogeneities; this however has two major difficulties: first, the general dynamics for effective LQG are not presently known beyond spherical symmetry, so considerable work will be required simply to derive the equations of motion; and second, more powerful analytic and numeric methods will be required, although recent progress on this problem in classical general relativity (see, e.g., \cite{Cook:2020oaj, Creminelli:2020zvc, Wang:2021hzv}) is encouraging.

It is already known that loop quantum cosmology can provide suitable initial conditions for the onset of inflation in a homogeneous cosmology \cite{Ashtekar:2009mm, Corichi:2010zp, Linsefors:2013cd, Bonga:2015xna, Gupt:2013swa}, and an inflationary era will isotropize the (homogeneous but anisotropic) Bianchi spacetimes \cite{Gupt:2013swa}---based on this work it seems likely that, in the presence of a suitable inflaton field, inflation (and isotropization) will occur at least in the homogenized regions. To verify and quantify this, it would be interesting to couple an inflaton field to this model to compute the size the homogeneous regions expand to by the end of inflation, for a given inflationary potential, and to what degree the remaining inhomogeneities in these regions are further smoothed out.

Finally, the homogenization that is found to occur in these models comes from an effective description that includes leading order quantum gravity effects for sharply-peaked states, but neglects the effect of quantum fluctuations. Since quantum gravity effects can homogenize the expectation values of such states in this approximation, it is natural to ask what may happen when quantum fluctuations are included. Do quantum fluctuations affect (either enhancing or destroying) the smoothing mechanism observed here? And this also raises the complementary question of whether quantum gravity could also `smooth out' and dynamically minimize quantum fluctuations (at least in some regions of space).  In other words, can quantum gravity dynamically drive the wave function of the universe to a state that is well approximated (at least in some region of space) by the tensor product of a homogeneous background and something close to an appropriate vacuum state for linear perturbations? These questions are left for future work.

\acknowledgments

\noindent
I thank Viqar Husain for helpful discussions.
This work was supported in part by the Natural Sciences and Engineering Research Council of Canada.

\raggedright

\end{document}